\documentclass[aps,prd,onecolumn,amsmath,11pt,superscriptaddress,floatfix,nofootinbib,preprintnumbers]{revtex4-2}

\usepackage[english]{babel}
\usepackage{xcolor}

\usepackage{amsmath,amstext,amsthm,bm,bbm,slashed,cancel,latexsym,mathtools,amssymb}
\usepackage{braket}
\usepackage{graphicx}
\usepackage{epstopdf}
\usepackage{keyval}
\usepackage{subfigure}
\usepackage{afterpage}
\usepackage{lipsum}
\usepackage{soul}
\usepackage{simplewick}
\usepackage{simpler-wick}
\usepackage[normalem]{ulem}

\usepackage{tikz}

\usepackage{calligra}
\usepackage[T1]{fontenc}
\usepackage{newunicodechar}
\newunicodechar{γ}{\ensuremath{\gamma}}
\usepackage{hyperref}
\hypersetup{colorlinks, linkcolor = [rgb]{0,0,0.5}, citecolor = [rgb]{0,0,0.5}, urlcolor = [rgb]{0,0,0.5}}
\usepackage{bookmark}
\allowdisplaybreaks[4]

\newcommand*{\SDU}{Key Laboratory of Particle Physics and Particle Irradiation (MOE), Institute of Frontier and Interdisciplinary Science, Shandong University, Qingdao, Shandong 266237, China}
\newcommand*{\SCNT}{Southern Center for Nuclear-Science Theory (SCNT), Institute of Modern Physics, Chinese Academy of Sciences, Huizhou 516000, China}

\begin{document}

\title{Azimuthal Correlations of Heavy-Quark Pairs in Ultraperipheral Collisions}

\author{Qiong Huang}\affiliation{\SDU}
\author{Jing Zhao}\affiliation{\SDU}
\author{Tianbo Liu}\email{liutb@sdu.edu.cn}\affiliation{\SDU}\affiliation{\SCNT}
\author{Ya-jin Zhou}\email{zhouyj@sdu.edu.cn}\affiliation{\SDU}

\begin{abstract}
We study azimuthal correlations of $D\bar D$ and $B\bar B$ pairs in Xn0n-selected Pb--Pb ultraperipheral collisions at $\sqrt{s}=5.36~\mathrm{TeV}$. The selection requires neutron emission from the target nucleus and no neutron emission from the photon source. The unpolarized and linearly polarized Weizs\"acker--Williams gluon transverse-momentum-dependent distributions are modeled in the Color Glass Condensate (CGC) framework and evolved to the hard scale. We keep the full mass dependence of the Born coefficients and the one-gluon soft current. The photon flux includes the neutron selection and is averaged over the target area. We present the transverse-momentum and rapidity dependence of the $\cos(2\phi_h)$ and $\cos(4\phi_h)$ asymmetries. Over the ranges studied, both moments increase with the pair total transverse momentum and decrease with its relative transverse momentum. For the lowest relative transverse momentum interval considered, the $D\bar D$ moments are larger than those for $B\bar B$. In the kinematic region considered, the $\cos(2\phi_h)$ asymmetry is much larger than the $\cos(4\phi_h)$ asymmetry, and the contribution from gluon linear polarization is small compared with the total angular modulation.
\end{abstract}


\maketitle

\section{Introduction}
Ultraperipheral collisions (UPCs) of relativistic heavy ions provide a clean environment to study photon-induced reactions at high energies~\cite{Bertulani_2005,Baltz:2007kq}. In these events, the impact parameter exceeds the sum of the nuclear radii, and direct hadronic interactions between the two ions are suppressed. Within the equivalent photon approximation (EPA), the electromagnetic field of a nucleus is described as a flux of quasi-real photons~\cite{vonWeizsacker:1934nji,Williams:1934ad}. These photons scatter off the other nucleus and can probe its gluon structure at small $x$. Recently, the CMS Collaboration measured inclusive $D^0$ photoproduction in Pb--Pb UPCs, using neutron emissions and a rapidity gap to select photonuclear events~\cite{CMS:2026D0}. This measurement provides an experimental basis for further studies of heavy-flavor production and pair correlations in neutron-selected UPCs.

Heavy-quark pair photoproduction is sensitive to transverse-momentum-dependent (TMD) gluon distributions. The quark mass provides an additional perturbative scale, while the nearly back-to-back configuration gives access to the transverse motion of gluons in the target. In the correlation limit, the total transverse momentum of the pair is much smaller than its relative transverse momentum. In this limit, the cross section is sensitive to the unpolarized and linearly polarized Weizs\"acker--Williams (WW) gluon TMDs~\cite{Boer_2011,Pisano_2013}. At small $x$, this TMD description can also be obtained from the Color Glass Condensate (CGC) calculation~\cite{DominguezXiaoYuan:2011,Dominguez_2011}. The linearly polarized WW distribution in an unpolarized nucleus can generate a $\cos(2\phi)$ modulation, where $\phi$ is the angle between the total and relative transverse momenta of the pair~\cite{PhysRevD.84.051503,PhysRevD.85.045003}.

Soft-gluon radiation also affects the pair transverse momentum and its azimuthal dependence. It can generate angular asymmetries even in the absence of linearly polarized gluons~\cite{PhysRevLett.126.142001,PhysRevD.104.054037}. This contribution must therefore be included when studying gluon polarization through pair correlations. In the back-to-back region, large logarithms of the ratio between the hard and transverse-momentum scales require TMD evolution and Sudakov resummation, which can be combined with the small-$x$ description~\cite{Mueller:2013wwa}. The heavy-quark mass suppresses collinear radiation and enters the soft factor~\cite{ZhuSunYuan:2013,Zhang:2017HQ}. Keeping this mass dependence is relevant for bottom production, where the mass need not be small compared with the quark transverse momentum.

In Ref.~\cite{marquet2025unifiedresummationsoftgluon}, the authors have studied heavy-meson pair photoproduction in a framework combining WW gluon distributions, mass effects, soft-gluon resummation, and fragmentation. To allow comparison with experimental measurements in the Xn0n event class, we include the corresponding neutron selection in the photon flux in this paper. The Xn0n class requires neutron emission on one side and no neutron emission on the other. Additional Coulomb excitation can change this classification, and the photon flux varies across the transverse area of the target~\cite{GimenoEstivill:2025D0,KleinNystrand:1999}. The neutron selection and the average over the target area thus affect the photon-flux weight entering the integrations over the final-state kinematics. 

In this work, we calculate azimuthal correlations of $D\bar D$ and $B\bar B$ pairs in Xn0n-selected Pb--Pb UPCs at $\sqrt{s}=5.36~\mathrm{TeV}$. We combine the conditional photon flux averaged over the target area with WW gluon TMDs and their scale evolution, taking the gluon distributions to be uniform inside the target. We keep the full mass dependence of the Born coefficients and the one-gluon soft current, and use Peterson fragmentation functions to obtain the meson cross sections. We present the $\cos(2\phi_h)$ and $\cos(4\phi_h)$ angular asymmetries and their transverse-momentum and rapidity dependence. We also examine the contribution of gluon linear polarization. For the kinematics tested, this contribution is smaller than the combined effect of soft radiation and fragmentation.

The paper is organized as follows. In Sec.~\ref{sec:framework_v2}, we present the theoretical framework for heavy-quark pair production in UPCs. Sec.~\ref{sec:numerical_v2} describes heavy-quark fragmentation and presents numerical results for the azimuthal asymmetries of heavy-meson pairs. Finally, we summarize our conclusions in Sec.~\ref{sec:conclusions_v2}. Details of the finite-mass soft current and the running-coupling prescription are given in Appendices~\ref{app:soft_mass_v1} and~\ref{app:evolution_numerics_v1}.

\section{Theoretical framework}
\label{sec:framework_v2}

In this section, we derive the differential cross section for heavy quark pair production in UPCs,
\begin{equation}
    \begin{gathered}
    {A_1+A_2\to A_1'+Q(k_1)+\bar Q(k_2)+X.}
    \end{gathered}
\end{equation}
The quantities in parentheses denote four-momenta. The recoiling photon source is denoted by $A_1'$, and $X$ denotes the unobserved final state. We use $P_1$ and $P_2$ for the momenta per nucleon of the incoming nuclei, with $s=(P_1+P_2)^2$. The momentum fractions $x_\gamma$ and $x_g$ are defined relative to $P_1$ and $P_2$, respectively.

In UPCs, the impact parameter of the nuclear collision satisfies $\tilde{b}_\perp\geq R_1+R_2$, where $R_1$ and $R_2$ are the radii of the two nuclei. The two nuclei do not overlap, and direct hadronic interactions between them are suppressed. The electromagnetic field of a relativistic nucleus can be described as a flux of quasi-real photons in the equivalent photon approximation (EPA) \cite{Bertulani_2005,Baltz:2007kq}. We take $A_1$ as the photon source and $A_2$ as the target. A photon from $A_1$ interacts with a gluon in $A_2$ to produce the heavy-quark pair. We keep the impact-parameter dependence of the photon flux $f_\gamma(x_\gamma,\rho_\perp)$, where $\rho_\perp$ is the distance from the center of $A_1$ to the hard interaction point, $x_\gamma=2\omega/\sqrt{s}$, and $\omega$ is the photon energy. For a nucleus with charge $Z$, the flux of transversely polarized photons is given by \cite{Bertulani_2005,Baltz:2007kq,Schafer:2020bnm}
\begin{equation}
    \begin{gathered}
    {
    x_\gamma f_\gamma(x_\gamma,\rho_\perp)
    =\frac{Z^2\alpha_e}{\pi^2}
    \frac{\omega^2}{\gamma_L^2}
    K_1^2\!\left(\frac{\omega\rho_\perp}{\gamma_L}\right),}
    \end{gathered}
    \label{eq:epa_v1}
\end{equation}
where \(K_1\) is the modified Bessel function of the second kind and \(\gamma_L\) is the Lorentz boost factor of the beam nucleus. For an ultrarelativistic ion beam,
\begin{equation}
    \omega = \frac{x_\gamma \sqrt{s}}{2},
    \qquad
    \gamma_L = \frac{\sqrt{s}}{2m_N},
\end{equation}
where \(m_N\) denotes the nucleon mass.

 In this work we consider Pb--Pb collisions, for which \(Z=82\), \(A=208\), and \(R_1=R_2\equiv R_A\).

\subsection[Conditional Neutron Selection and Photon Flux]{Conditional Neutron Selection and Photon Flux}

The experimental Xn0n selection requires at least one neutron from the target $A_2$ and no neutrons from the photon source $A_1$. These neutrons are measured by zero-degree calorimeters~\cite{CMS:2026D0}. In the inclusive channel considered here, we assume that the hard photonuclear interaction breaks up $A_2$ and satisfies the target-side Xn condition \cite{GimenoEstivill:2025D0}. Additional soft photon exchanges between the nuclei can cause Coulomb dissociation even when $\tilde{b}_\perp\geq 2R_A$. Neutron emission from $A_1$ would then remove the event from the selected sample, while extra neutron emission from $A_2$ leaves the Xn condition unchanged. Thus only the no-neutron probability of $A_1$ is needed as an additional electromagnetic weight. Treating the hard process and Coulomb dissociation as independent at fixed $\tilde{b}_\perp$, we write \cite{PhysRevE.54.4233,Baltz_1998}
\begin{equation}
    \begin{gathered}
    {P(\tilde{b}_\perp)=\exp\!\left(-\frac{S}{\tilde{b}_\perp^2}\right).}
    \end{gathered}
    \label{eq:neutron_v1}
\end{equation}
In the numerical calculation we use $S=(17.4~{\rm fm})^2$ \cite{Baltz_1998,PhysRevE.54.4233}.
Since the hard interaction already supplies the target-side neutrons, we do not multiply this probability by an additional target excitation factor.

\begin{figure}[htbp]
    \centering
    \includegraphics[width=\textwidth]{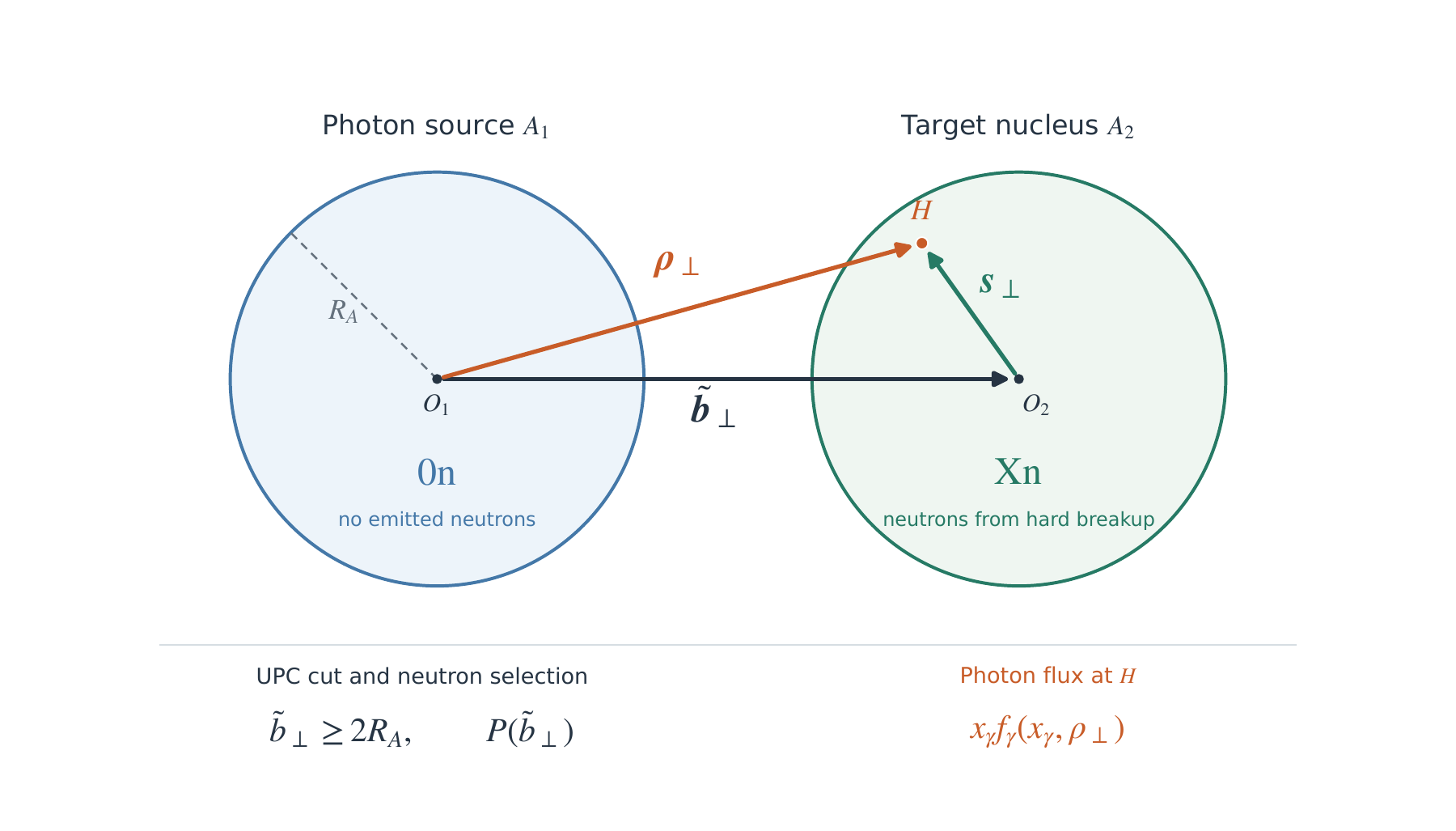}
    \caption{Transverse geometry of the Xn0n-selected UPC process. The hard interaction point $H$ is at $\bm s_\perp$ relative to the target center and at $\bm\rho_\perp=\tilde{\bm b}_\perp+\bm s_\perp$ relative to the photon source.}
    \label{fig:upc_geometry_v1}
\end{figure}

The photon flux also varies across the target and must be averaged over its transverse area~\cite{KleinNystrand:1999}. The transverse geometry is shown in Fig.~\ref{fig:upc_geometry_v1}. Let the impact-parameter vector $\tilde{\bm b}_\perp$ point from the center of $A_1$ to the center of $A_2$, and let $\bm s_\perp$ be the hard interaction position measured from the center of $A_2$. The impact parameter entering the photon flux is then
\begin{equation}
    \rho_\perp=\left|\tilde{\bm b}_\perp+\bm s_\perp\right|.
    \label{eq:rho_v1}
\end{equation}
We take $R_A=6.4~\mathrm{fm}$ and impose the UPC cut $\tilde{b}_\perp\geq 2R_A$. The impact parameter $\tilde{\bm b}_\perp$ and the position $\bm s_\perp$ are distinct from the Fourier coordinate $\bm b_\perp$ used below for the gluon TMDs.

We assume that the unpolarized and linearly polarized gluon distributions per unit transverse area are uniform inside the target. Their common spatial weight is
\begin{equation}
    T_{\rm disk}(s_\perp)=\frac{\Theta(R_A-s_\perp)}{\pi R_A^2},
    \qquad
    \int d^2\bm s_\perp\,T_{\rm disk}(s_\perp)=1.
    \label{eq:disk_v1}
\end{equation}
Both gluon distributions use the same position-independent saturation scale $Q_{s,A}(x)$. The gluon TMDs used below already contain the nuclear area $S_\perp=\pi R_A^2$, while $T_{\rm disk}$ is normalized to unity. Averaging the photon flux over this distribution and imposing the neutron selection gives
\begin{equation}
    \Phi_{\mathrm{Xn0n}}(x_\gamma)
    =\int_{\tilde{b}_\perp\geq 2R_A}d^2\tilde{\bm b}_\perp\,
    P(\tilde{b}_\perp)
    \int d^2\bm s_\perp\,T_{\rm disk}(s_\perp)\,
    x_\gamma f_\gamma(x_\gamma,\rho_\perp).
    \label{eq:conditional_flux_v1}
\end{equation}
Here the neutron probability depends on the impact parameter $\tilde b_\perp$, whereas the photon flux depends on $\rho_\perp$, measured from the photon source to the hard interaction point.

The differential cross section can then be written as
\begin{equation}
    \frac{\mathrm{d}\sigma^{A_1A_2\to A_1' Q\bar Q X}_{\mathrm{Xn0n}}}
    {\mathrm{d}^2\bm{k}_{1\perp}\,\mathrm{d}^2\bm{k}_{2\perp}\,\mathrm{d}y_1\,\mathrm{d}y_2}
    =
    \Phi_{\mathrm{Xn0n}}(x_\gamma)\,
    \frac{\mathrm{d}\sigma^{\gamma {A_2}\to Q\bar Q X}}
    {\mathrm{d}^2\bm{k}_{1\perp}\,\mathrm{d}^2\bm{k}_{2\perp}\,\mathrm{d}y_1\,\mathrm{d}y_2}.
    \label{eq:upc_cross_section_v1}
\end{equation}
Here $\bm k_{1\perp}$ and $\bm k_{2\perp}$ are the transverse momenta of the heavy quark and antiquark, and $y_1$ and $y_2$ are their rapidities. The $x_\gamma$ dependence of $\Phi_{\mathrm{Xn0n}}$ is kept inside the integrations over the final-state kinematics. This expression refers to photon emission from $A_1$; the opposite direction is obtained by interchanging the two nuclei and applying the corresponding neutron selection.
Here $\mathrm{d}\sigma^{\gamma {A_2}\to Q\bar Q X}$ denotes the differential cross section for the photon--nucleus subprocess, which will be discussed in the next subsection.

\subsection{WW gluon TMD distribution and small-x evolution}
\label{Sec. 2.2}

The photonuclear process considered in this work is
\begin{equation}
    \begin{gathered}
    {\gamma(l)+{A_2}\to Q(k_1)+\bar Q(k_2)+X.}
    \end{gathered}
\end{equation}
We neglect the photon transverse momentum and take $l\simeq x_\gamma P_1$. We choose $P_1$ along the positive longitudinal direction and $P_2$ along the negative direction.

In the TMD factorization framework, the differential cross section for heavy-quark pair production can be written as~\cite{Boer_2011,Pisano_2013}
\begin{equation}
    \begin{aligned}
        \frac{\mathrm d \sigma^{\gamma {A_2} \to Q\bar Q X}}
        {\mathrm d^2 \bm k_{1\perp}\,\mathrm d^2 \bm k_{2\perp}\,\mathrm d y_1\,\mathrm d y_2}
        &={\left[
        -\frac{1}{2}g_\perp^{\beta\alpha}\,
        x_g f_1^g(x_g,p_\perp)
        +\left(\frac{p_\perp^\beta p_\perp^\alpha}{\bm p_\perp^2}
        +\frac{1}{2}g_\perp^{\beta\alpha}\right)
        x_g h_1^{\perp g}(x_g,p_\perp)\right]H_{\alpha\beta}\, ,}
    \end{aligned}
    \label{eq:tmd_tensor_v1}
\end{equation}
where $f_1^g$ and $h_1^{\perp g}$ denote the unpolarized and linearly polarized gluon TMDs, respectively, and $H_{\alpha\beta}$ is the perturbatively calculable hard factor. The incoming gluon carries transverse momentum $\bm p_\perp$. Here $H_{\alpha\beta}$ includes the incident flux, phase-space normalization, and photon polarization average for the Born subprocess $\gamma g\to Q\bar Q$, with the heavy-quark mass included. The photon transverse momentum is neglected in this hard scattering expression.

We define $\bm q_\perp=\bm k_{1\perp}+\bm k_{2\perp}$ and $\bm P_\perp=(\bm k_{1\perp}-\bm k_{2\perp})/2$, and denote the angle between them by $\phi$. At Born level, neglecting the photon transverse momentum, the pair imbalance equals the transverse momentum of the incoming gluon, i.e., $\bm q_\perp=\bm p_\perp$. The unpolarized distribution gives the azimuthally independent contribution, while the linearly polarized distribution generates a $\cos(2\phi)$ modulation \cite{Dominguez_2011,Boer_2011}. Additional soft radiation changes the pair imbalance and also produces angular modulations~\cite{PhysRevLett.126.142001,PhysRevD.104.054037}, as discussed later in Sec.~\ref{subsec:final_state_resummation}.

For an unpolarized hadron with momentum $P$ chosen along the positive longitudinal direction, the WW gluon correlator with future-pointing gauge links is defined as \cite{PhysRevD.63.094021,PhysRevD.84.051503}
\begin{equation}
\begin{gathered}
{\begin{aligned}
&\int\frac{\mathrm d y^-\,\mathrm d^2\bm y_\perp}{(2\pi)^3 P^+}
 e^{ixP^+y^--i\bm k_\perp\cdot\bm y_\perp}
 \langle P|F_a^{+i}(0)\,\mathcal U^{[+]}_{ab}(0,y)\,F_b^{+j}(y)|P\rangle\big|_{y^+=0}\\
&\qquad=-\frac{g_\perp^{ij}}2\,xf_1^g(x,k_\perp)
 +\left(\frac{k_\perp^ik_\perp^j}{\bm k_\perp^2}+\frac{g_\perp^{ij}}2\right)xh_1^{\perp g}(x,k_\perp).
\end{aligned}}
\end{gathered}
\label{eq:ww_correlator_v1}
\end{equation}
Here $x=k^+/P^+$, and \mbox{\(\mathcal U^{[+]}_{ab}(0,y)=[L_0^\dagger L_y]_{ab}\)}. Both $L_0$ and $L_y$ run to $+\infty^-$ in the adjoint representation, with the transverse connection at infinity understood \cite{DominguezXiaoYuan:2011,Dominguez_2011,PhysRevD.84.051503}. At small $x$, the WW distributions in the McLerran--Venugopalan (MV) model can be written in the large-$N_c$ and leading-logarithmic approximation as \cite{Dominguez_2011,PhysRevD.84.051503}
\begin{equation}
\begin{aligned}
x f_{1,\mathrm{WW}}^g(x,k_\perp)
=
\frac{N_c S_\perp}{2\pi^3 \alpha_s}
\int_{{0}}^{{\infty}} \mathrm d b_\perp \, b_\perp
J_0(b_\perp k_\perp)
\left[
1-\exp\left(
-\frac{Q_{s,A}^2(x) b_\perp^2}{4}
\ln \frac{1}{b_\perp^2 \Lambda_{\mathrm{QCD}}^2}
\right)
\right]
\frac{1}{b_\perp^2}\, ,
\end{aligned}
\label{eq:ww_f_v1}
\end{equation}
\begin{equation}
\begin{aligned}
x h_{1,\mathrm{WW}}^{\perp g}(x,k_\perp)
=
\frac{N_c S_\perp}{2\pi^3 \alpha_s}
\int_{{0}}^{{\infty}} \mathrm d b_\perp \, b_\perp
J_2(b_\perp k_\perp)
\left[
1-\exp\left(
-\frac{Q_{s,A}^2(x) b_\perp^2}{4}
\ln \frac{1}{b_\perp^2 \Lambda_{\mathrm{QCD}}^2}
\right)
\right]
\frac{1}{b_\perp^2 \ln \frac{1}{b_\perp^2 \Lambda_{\mathrm{QCD}}^2}} \, ,
\end{aligned}
\label{eq:ww_h_v1}
\end{equation}
where $S_\perp=\pi R_A^2$ is the transverse area of the target nucleus. The coupling $\alpha_s$ in the overall normalization of Eqs.~\eqref{eq:ww_f_v1} and \eqref{eq:ww_h_v1} is treated as a constant, independently of the running coupling used for evolution. The common normalization cancels in the normalized azimuthal asymmetries.
 As an infrared prescription for the WW model, we regulate the logarithm according to $\ln \frac{1}{b_\perp^2 \Lambda_{\mathrm{QCD}}^2}\;\to\;\ln \left(\frac{1}{b_\perp^2 \Lambda_{\mathrm{QCD}}^2} + e\right)$~\cite{PhysRevD.94.014030}. This replacement is applied both in the exponent and in the denominator of the polarized distribution.

We determine the $x$ dependence of the scale entering these WW distributions from the proton dipole amplitude. We first solve the impact-parameter-independent running-coupling Balitsky--Kovchegov (rcBK) equation~\cite{Balitsky:1996,Kovchegov:1999} for
\begin{equation}
    N_p(\bm r, Y) = 1 - S_p(\bm r, Y) \, ,
\end{equation}
where
\begin{equation}
    S_p(\bm r, Y)
    =
    \frac{1}{N_c}
    \left\langle
    \mathrm{Tr}\,
    U(\bm x)\, U^\dagger(\bm y)
    \right\rangle_Y ,
    \qquad
    \bm r = \bm x - \bm y .
\end{equation}
Here $U$ is a Wilson line in the fundamental representation, and the average is over the target color fields. The rapidity variable is defined by $Y = \ln (x_0/x)$. The evolution equation reads
\begin{equation}
    \begin{aligned}
        \frac{\partial N_p(r,{Y})}{\partial Y}=\int\mathrm{d}{{}^2}\bm{r}_1K^{\text{run}}(\bm{r},\bm{r}_1,\bm{r}_2)[N_p(r_1,Y)+N_p(r_2,Y)-N_p(r,Y)-N_p(r_1,Y)N_p(r_2,Y)],
    \end{aligned}
\end{equation}
where $\bm r_2=\bm r-\bm r_1$ and $r_i=|\bm r_i|$. We use Balitsky's running-coupling kernel~\cite{Balitsky:2006wa},
\begin{equation}
    \begin{aligned}
        K^{\text{run}}(\bm{r},\bm{r}_1,\bm{r}_2)=\frac{N_c\alpha_s(r^2)}{2\pi^2}\left[\frac{r^2}{r_1^2r_2^2}+\frac{1}{r_1^2}\left(\frac{\alpha_s(r_1^2)}{\alpha_s(r_2^2)}-1\right)+\frac{1}{r_2^2}\left(\frac{\alpha_s(r_2^2)}{\alpha_s(r_1^2)}-1\right)\right],
    \end{aligned}
\end{equation}
with the one-loop running coupling
\begin{equation}
    \begin{aligned}
        \alpha_s(r{{}^2})=\frac{12\pi}{(33-2n_f)\log\left(\frac{4C^2}{r^2\Lambda_{\mathrm{QCD}}^2}\right)}.
    \end{aligned}
\end{equation}
In practice, we solve this equation numerically using the MV initial condition for the proton dipole amplitude at $x_0=0.01$ ($Y=0$) \cite{PhysRevD.49.2233,PhysRevD.88.114020}
\begin{equation}
    \begin{gathered}
    {N_p(r,0)=1-\exp\left[-\frac{r^2Q_{0,p}^2}{4}\ln\left(\frac1{r\Lambda_{\mathrm{QCD}}}+e\right)\right].}
    \end{gathered}
    \label{eq:mv_initial_v1}
\end{equation}

\begin{figure}[htpb]
    \centering
    \includegraphics[scale=0.9]{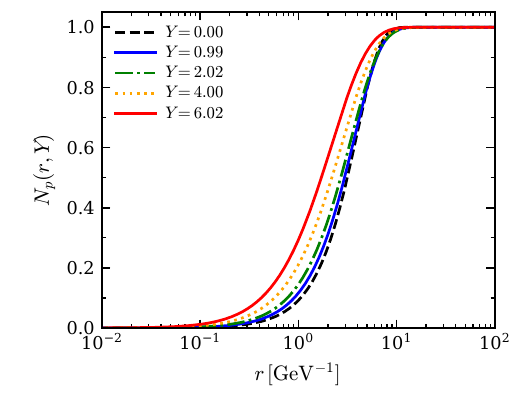}%
    \hfill
    \includegraphics[scale=0.9]{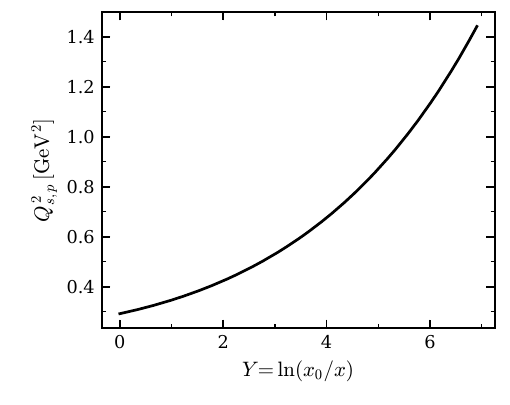}
    \caption{
    Small-$x$ evolution of the proton dipole amplitude $N_p(r,Y)$ (left) and the extracted saturation scale $Q_{s,p}^2(Y)$ (right). }
    \label{Fig. small x evo}
\end{figure}

We use $Q_{0,p}^2=0.2~\mathrm{GeV}^2$ and $\Lambda_{\mathrm{QCD}}=0.241~\mathrm{GeV}$ as in Ref.~\cite{Zhao_2021}, with $N_c=3$ and $n_f=3$. We choose $C^2=14.5$ from the MV fit in Ref.~\cite{PhysRevD.88.114020}. The running coupling is frozen at $0.7$ for all distances beyond the point where it first reaches this value~\cite{Albacete_2011}. At each $Y$, we extract the proton saturation scale from \cite{PhysRevD.88.114020}
\begin{equation}
    N_p(r_s(Y),Y)=1-e^{-1/2},\qquad Q_{s,p}^2(Y)=\frac{2}{r_s^2(Y)}.
\end{equation}
The initial-condition coefficient $Q_{0,p}^2$ differs from the extracted saturation scale. With these inputs, $Q_{s,p}^2(x_0)\simeq\,0.292~\mathrm{GeV}^2$. Figure~\ref{Fig. small x evo} shows the evolved proton amplitude and the extracted scale, with $x=x_0e^{-Y}$.

For the WW input we adopt the nuclear enhancement prescription \cite{Zhao_2021}
\begin{equation}
    Q_{s,A}^2(x)=\frac{C_A}{C_F}\,A^{1/3}Q_{s,p}^2(x).\qquad
\end{equation}
The factor $A^{1/3}$ accounts for the nuclear enhancement, and $C_A/C_F$ introduces the adjoint-to-fundamental color-factor ratio. Here $Q_{s,A}^2$ is an effective scale parameter in the WW model.

\subsection{TMD evolution}
\label{Sec. 2.3}

When the measured transverse momentum is much smaller than the hard scale $Q$, soft and collinear radiation produces large logarithms that spoil a fixed-order expansion. We resum the logarithms associated with the incoming gluon using the Collins--Soper--Sterman (CSS) formalism~\cite{CollinsSoper:1981,CollinsSoperSterman:1985}, which accounts for the resulting transverse-momentum broadening. Its application to small-$x$ distributions was developed in Ref.~\cite{MuellerXiaoYuan:2013PRL}. For heavy-quark pair production, we choose $Q^2=(k_1+k_2)^2$. The evolved WW distributions are written as \cite{PhysRevD.105.096017}
\begin{equation}
\begin{gathered}
{
x f_{1,\mathrm{WW}}^g(x,k_\perp;Q)
=\int_0^\infty\frac{\mathrm d b_\perp\,b_\perp}{2\pi}\,
J_0(k_\perp b_\perp)\,e^{-S_{\rm ini}(b_\perp,Q)}
x\tilde f_{1,\mathrm{WW}}^g(x,b_\perp^*),}
\end{gathered}
\label{eq:ww_evolved_f_v1}
\end{equation}
\begin{equation}
\begin{gathered}
{
x h_{1,\mathrm{WW}}^{\perp g}(x,k_\perp;Q)
=\int_0^\infty\frac{\mathrm d b_\perp\,b_\perp}{2\pi}\,
J_2(k_\perp b_\perp)\,e^{-S_{\rm ini}(b_\perp,Q)}
x\tilde h_{1,\mathrm{WW}}^{\perp g}(x,b_\perp^*).}
\end{gathered}
\label{eq:ww_evolved_h_v1}
\end{equation}
The coordinate-space inputs are defined from Eqs.~\eqref{eq:ww_f_v1} and \eqref{eq:ww_h_v1} by
\begin{align}
x\tilde f_{1,\mathrm{WW}}^g(x,b_\perp)
&=2\pi\int_0^\infty\mathrm d k_\perp\,k_\perp\,
J_0(k_\perp b_\perp)\,x f_{1,\mathrm{WW}}^g(x,k_\perp),\\
x\tilde h_{1,\mathrm{WW}}^{\perp g}(x,b_\perp)
&=2\pi\int_0^\infty\mathrm d k_\perp\,k_\perp\,
J_2(k_\perp b_\perp)\,x h_{1,\mathrm{WW}}^{\perp g}(x,k_\perp).
\end{align}
The $J_2$ transform follows from the rank-two tensor structure of the linearly polarized distribution. 

Here the Sudakov factor is decomposed as
\begin{equation}
\begin{gathered}
{
S_{\rm ini}(b_\perp,Q)
=S_A\!\left((\mu'_{b_*})^2,Q^2\right)+S_{\rm NP}(b_\perp^2,Q^2),}
\end{gathered}
\label{eq:ww_sini_v1}
\end{equation}
with perturbative part~\cite{Boer:2017TMD}
\begin{equation}
\begin{gathered}
{
S_A\!\left((\mu'_{b_*})^2,Q^2\right)
=\frac{C_A}{2\pi}\int_{(\mu'_{b_*})^2}^{Q^2}
\frac{\mathrm d\mu^2}{\mu^2}\,\alpha_s(\mu)
\left[\ln\frac{Q^2}{\mu^2}-\frac{11-2n_f/C_A}{6}\right].}
\end{gathered}
\label{eq:ww_sa_v1}
\end{equation}

We use the $b_*$ prescription and the regulated lower scale \cite{Boer:2014tka,Boer:2017TMD}
\begin{equation}
 b_\perp^*=\frac{b_\perp}{\sqrt{1+b_\perp^2/b_{\max}^2}},\qquad
 \mu_{b_*}^2=\frac{4e^{-2\gamma_E}}{(b_\perp^*)^2},\qquad
 (\mu'_{b_*})^2=\left(\frac{1}{\mu_{b_*}^2}+\frac{1}{Q^2}\right)^{-1}.
\end{equation}
Here $\gamma_E$ is Euler's constant. The $b_*$ prescription limits the large-distance input, while $S_{\rm NP}$ describes the remaining long-distance dependence. The regulated scale approaches $Q$ as $b_\perp\to0$. We use one-loop running for $\alpha_s(\mu)$ with $n_f=3$ and $\Lambda_{\mathrm{QCD}}=0.241~\mathrm{GeV}$. 

For the nonperturbative Sudakov factor, we adopt the parametrization of Ref.~\cite{PhysRevD.83.114042}, with the color-factor scaling for gluons~\cite{Boer:2014tka,Boer:2017TMD},
\begin{equation}
\begin{gathered}
{
S_{\rm NP}(b_\perp^2,Q^2)
=\frac{C_A}{2C_F}\left[g_1+g_2\ln\frac{Q}{2Q_0}
+2g_1g_3\ln\frac{10x x_{\rm NP}}{x_{\rm NP}+x}\right]b_\perp^2,}
\end{gathered}
\label{eq:ww_snp_v1}
\end{equation}
with $b_{\max}=1.5~\mathrm{GeV}^{-1}$, $g_1=0.201~\mathrm{GeV}^2$, $g_2=0.184~\mathrm{GeV}^2$, $g_3=-0.129$, $x_{\rm NP}=0.009$, and $Q_0=1.6~\mathrm{GeV}$~\cite{KonychevNadolsky:2006,PhysRevD.83.114042}. We use the same $S_{\rm NP}$ for the unpolarized and linearly polarized distributions.

\begin{figure}[htbp]
    \centering
    \includegraphics[width=\textwidth]{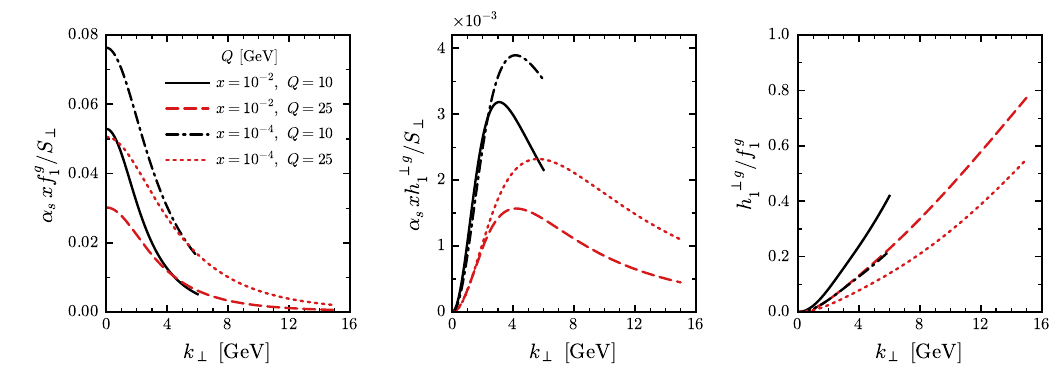}
    \caption{
    Evolved WW gluon distributions $\alpha_s x f_1^g/S_\perp$ (left), $\alpha_s x h_1^{\perp g}/S_\perp$ (middle), and their ratio $h_1^{\perp g}/f_1^g$ (right) for Pb. Black and red curves correspond to $Q=10$ and $25~\mathrm{GeV}$, respectively; the four line styles distinguish the $(x,Q)$ combinations listed in the left panel. }
    \label{Fig. gluon TMDs}
\end{figure}

The evolved WW distributions and their ratio are shown in Fig.~\ref{Fig. gluon TMDs}. As $Q$ increases, both distributions become broader, and the peak of the linearly polarized distribution moves to larger $k_\perp$. The ratio $h_1^{\perp g}/f_1^g$ increases with $k_\perp$, while at fixed $k_\perp$ it is smaller for larger $Q$ or smaller $x$. Thus, the relative size of the linearly polarized distribution is reduced in the low-transverse-momentum region. The curves at large $k_\perp$ illustrate the behavior of the model.

 \subsection{Final-State Soft Gluon Resummation}
 \label{subsec:final_state_resummation}

Radiation from the outgoing heavy quarks changes both the transverse-momentum distribution and its azimuthal dependence~\cite{CataniGrazziniTorre:2015,PhysRevLett.126.142001,PhysRevD.104.054037}. The quark mass suppresses collinear radiation~\cite{DokshitzerKhozeTroyan:1991}. We keep the full mass dependence of the one-gluon soft current \cite{ZhuSunYuan:2013,Zhang:2017HQ}, in addition to the incoming-gluon evolution of Sec.~\ref{Sec. 2.3}. The partonic process is
\begin{equation}
    \begin{gathered}
    {\gamma(l)+g(x_g P_2+p_\perp)\to Q(k_1)+\bar Q(k_2)+X.}
    \end{gathered}
\end{equation}

At one loop, the real-emission contribution has the form
\begin{equation}
    \frac{\,{d\sigma_{\rm real}^{(1)}}}
    {d^2\bm P_\perp d^2\bm q_\perp}
    =
    \int d^2\bm q'_\perp d^2\bm k_{g\perp}
    \frac{d\sigma_{\rm LO}}
    {d^2\bm P_\perp d^2\bm q'_\perp}
    S_r(\bm k_{g\perp})
    \delta^{(2)}
    \left(
        \bm q_\perp-\bm q'_\perp{+}\bm k_{g\perp}
    \right),
\end{equation}
where $\bm q'_\perp=\bm q_\perp+\bm k_{g\perp}$ is the total transverse momentum entering the hard subprocess, with $\bm k_{g\perp}$ the transverse momentum of the emitted soft gluon. The variables $\bm q_\perp$ and $\bm P_\perp$ are defined in Sec.~\ref{Sec. 2.2}. For an emitted gluon of rapidity $y_g$ and angle $\phi_{k_gP}=\angle(\bm k_{g\perp},\bm P_\perp)$, the real-emission soft factor $S_r(\bm k_{g\perp})$ is written as
\begin{align}
 S_r(\bm k_{g\perp})&=\frac{\alpha_s(\mu)}{4\pi^2 k_{g\perp}^2}\int\mathrm d y_g\,W(y_g,\phi_{k_gP}),\label{eq:real_soft_factor_v2}\\
 W(y_g,\phi_{k_gP})&=k_{g\perp}^2\Bigg\{
 C_A\left[\frac{p\cdot k_1}{(p\cdot k_g)(k_1\cdot k_g)}
          +\frac{p\cdot k_2}{(p\cdot k_g)(k_2\cdot k_g)}\right]\notag\\
 &\quad-(C_A-2C_F)\frac{k_1\cdot k_2}{(k_1\cdot k_g)(k_2\cdot k_g)}
 -C_F M_Q^2\left[\frac{1}{(k_1\cdot k_g)^2}+\frac{1}{(k_2\cdot k_g)^2}\right]\Bigg\}.
\label{eq:soft_current_mass_v1}
\end{align}
Here $M_Q$ is the heavy-quark mass. In the soft current, $p\simeq x_gP_2$ is the incoming gluon momentum in the collinear approximation. The last two terms are the massive self terms. The rapidity integral is regulated until the incoming contribution is subtracted.

The soft kernel is evaluated in a longitudinally boosted frame where the mean quark rapidity vanishes. With the incoming gluon along $-z$, the finite kernel and its angular moments are
\begin{align}
 K(\phi_{k_gP})&=\lim_{Y_{\rm cut}\to\infty}
 \left[\int_{-Y_{\rm cut}}^{Y_{\rm cut}}\mathrm d y_g\,W(y_g,\phi_{k_gP})-4C_A Y_{\rm cut}\right],\label{eq:soft_moments_mass_v1}\\
 K_{2n}&=\int_0^{2\pi}\frac{\mathrm d\phi_{k_gP}}{2\pi}\,
 K(\phi_{k_gP})\cos(2n\phi_{k_gP}),\qquad n=0,1,\ldots .
\label{eq:soft_harmonic_moments_v2}
\end{align}
The subtraction removes the incoming rapidity divergence. After the rapidity integration, quark--antiquark interchange leaves only even harmonics. The moments depend on $P_\perp$, $M_Q$, and $y_1-y_2$. The color factors, the explicit current and its harmonic expansion are given in Appendix~\ref{app:soft_mass_v1}.

 At logarithmic accuracy, the angular average and its virtual counterpart give the additional isotropic exponent~\cite{ZhuSunYuan:2013}

\begin{equation}
 S_{\rm soft}(b_\perp,Q,P_\perp,M_Q)
 =\int_{(\mu'_{b_*})^2}^{Q^2}\frac{\mathrm d\mu^2}{\mu^2}\,
 \frac{\alpha_s(\mu)}{2\pi}
 \left[C_A\ln\frac{P_\perp^2+M_Q^2}{M_Q^2}-2C_F
 -\frac{1+\beta^2}{2C_A\beta}\ln\frac{1+\beta}{1-\beta}\right],
\label{eq:soft_exponent_mass_v1}
\end{equation}
where $\beta=\sqrt{1-4M_Q^2/Q^2}$.

At leading power in the correlation limit, the pair invariant mass and longitudinal momentum fractions are
\begin{align}
 Q^2&=2(P_\perp^2+M_Q^2)\big[1+\cosh(y_1-y_2)\big],\label{eq:hard_kinematics_v1}\\
 x_\gamma&=\sqrt{\frac{P_\perp^2+M_Q^2}{s}}\left(e^{y_1}+e^{y_2}\right),\label{eq:hard_xgamma_v2}\\
 x_g&=\sqrt{\frac{P_\perp^2+M_Q^2}{s}}\left(e^{-y_1}+e^{-y_2}\right).
 \label{eq:hard_xg_v2}
\end{align}
 The lower scale $\mu'_{b_*}$ is the regulated scale defined in Sec.~\ref{Sec. 2.3}. We use one-loop running for the soft factor, continuous at $M_c=1.27~\mathrm{GeV}$ and $M_b=4.65~\mathrm{GeV}$, with $\Lambda_{n_f=3}=0.241~\mathrm{GeV}$.

For the nonzero harmonics, the virtual contribution is independent of $\phi_{k_gP}$ and does not contribute to the angular projection. We keep the angular soft factor to one loop \cite{PhysRevD.104.054037,marquet2025unifiedresummationsoftgluon}:
\begin{align}
 &1+\sum_{n=1}^{\infty}(-1)^n a_{2n}(b_\perp)\cos(2n\phi_{bP}),\label{eq:soft_angular_factor_v2}\\
 &a_{2n}(b_\perp)=\frac{\alpha_s(\mu'_{b_*})K_{2n}}{\pi}
 \int_0^Q\frac{\mathrm d k_{g\perp}}{k_{g\perp}}J_{2n}(k_{g\perp}b_\perp),
\label{eq:soft_angular_mass_v1}
\end{align}
where $\phi_{bP}=\angle(\bm b_\perp,\bm P_\perp)$. The Bessel function makes this integral finite at $k_{g\perp}=0$. We use $Q$ as the upper real-emission scale, so $a_{2n}$ vanishes as $b_\perp\to0$. The factor $(-1)^n$ follows from the angular Fourier transform.

Taking all these ingredients into account, the cross section for heavy-quark pair production can be written as
\begin{equation}
\begin{aligned}
 \frac{\mathrm d\sigma}{\mathrm d^2\bm q_\perp\,{\mathrm d^2\bm P_\perp\,\mathrm d y_1\,\mathrm d y_2}}
 &=\int\frac{\mathrm d^2\bm b_\perp}{(2\pi)^2}\,
 e^{i\bm q_\perp\cdot\bm b_\perp}
 e^{-S_{\rm ini}(b_\perp,Q)-S_{\rm soft}(b_\perp,Q,P_\perp,M_Q)}\\
 &\quad\times\left[\mathcal H_U\,x_g\tilde f_{1,\mathrm{WW}}^g(x_g,b_\perp^*)
 -\mathcal H_P\,x_g\tilde h_{1,\mathrm{WW}}^{\perp g}(x_g,b_\perp^*)\cos(2\phi_{bP})\right]\\
 &\quad\times\left[1+\sum_{n=1}^{\infty}(-1)^n a_{2n}(b_\perp)\cos(2n\phi_{bP})\right].
\end{aligned}
\label{eq:ww_soft_product_mass_v1}
\end{equation}
  The Born coefficients~\cite{Boer_2011,Pisano_2013}, together with the conditional photon flux, give
\begin{align}
 \mathcal H_U&=\frac{\alpha_e\alpha_s e_Q^2}{Q^4}
 \frac{(Q^2-2M_Q^2)M_Q^2+(Q^2-2P_\perp^2)P_\perp^2}{(M_Q^2+P_\perp^2)^2}
 \Phi_{\mathrm{Xn0n}}(x_\gamma),\label{eq:hard_coefficients_v1}\\
 \mathcal H_P&=\frac{\alpha_e\alpha_s e_Q^2}{Q^4}
 \frac{4M_Q^2P_\perp^2}{(M_Q^2+P_\perp^2)^2}
 \Phi_{\mathrm{Xn0n}}(x_\gamma).
 \label{eq:hard_coefficient_p_v2}
\end{align}
Here $e_Q$ is the quark electric charge in units of the positron charge. The minus sign of the polarized term follows from its rank-two Fourier transform. The resummed cross section can then be written as
\begin{equation}
 \frac{\mathrm d\sigma}
 {\mathrm d^2\bm q_\perp\,\mathrm d^2\bm P_\perp\,\mathrm d y_1\,\mathrm d y_2}
 =\mathcal C_0(q_\perp)
 +\mathcal C_2(q_\perp)\cos(2\phi)
 +\mathcal C_4(q_\perp)\cos(4\phi)+\cdots .
 \label{eq:partonic_harmonics_v2}
\end{equation}
The momentum-space coefficients are related to the coordinate-space coefficients by
\begin{equation}
 \mathcal C_{2n}(q_\perp)
 =(-1)^n\int_0^\infty\frac{b_\perp\,\mathrm d b_\perp}{2\pi}\,
 e^{-S_{\rm ini}-S_{\rm soft}}
 J_{2n}(b_\perp q_\perp)\,C_{2n}(b_\perp),
 \qquad n=0,1,\ldots ,
 \label{eq:harmonic_transform_v2}
\end{equation}
where
\begin{align}
 C_0(b_\perp)&=\mathcal H_U\,x_g\tilde f_{1,\mathrm{WW}}^g(x_g,b_\perp^*)
 +\frac{\mathcal H_P}{2}\,x_g\tilde h_{1,\mathrm{WW}}^{\perp g}(x_g,b_\perp^*)a_2,\label{eq:meson_coefficients_v1}\\
 C_2(b_\perp)&=-\mathcal H_U\,x_g\tilde f_{1,\mathrm{WW}}^g(x_g,b_\perp^*)a_2
 -\mathcal H_P\,x_g\tilde h_{1,\mathrm{WW}}^{\perp g}(x_g,b_\perp^*)\left(1+\frac{a_4}{2}\right),\label{eq:meson_coefficient_c2_v2}\\
 C_4(b_\perp)&=\mathcal H_U\,x_g\tilde f_{1,\mathrm{WW}}^g(x_g,b_\perp^*)a_4
 +\frac{\mathcal H_P}{2}\,x_g\tilde h_{1,\mathrm{WW}}^{\perp g}(x_g,b_\perp^*)(a_2+a_6).
 \label{eq:meson_coefficient_c4_v2}
\end{align}
The $a_6$ term in $C_4$ comes from the product of the polarized WW term and the sixth soft harmonic. Higher even harmonics from the same product are included in the numerical calculation.

\section{Numerical Analysis for Azimuthal Asymmetry}
\label{sec:numerical_v2}

We now include fragmentation to obtain the $D\bar D$ and $B\bar B$ cross sections. We denote the meson transverse momenta by $\bm p_{h1\perp}$ and $\bm p_{h2\perp}$, and define
\begin{equation}
 \bm P_{h\perp}=\frac{\bm p_{h1\perp}-\bm p_{h2\perp}}{2},
 \qquad \bm q_{h\perp}=\bm p_{h1\perp}+\bm p_{h2\perp},
 \qquad \phi_h=\angle(\bm q_{h\perp},\bm P_{h\perp}).
 \label{eq:meson_variables_v1}
\end{equation}
Here $\bm q_{h\perp}$ is the total transverse momentum of the meson pair, and $q_{h\perp}=|\bm q_{h\perp}|$.
In collinear fragmentation, we take $\bm p_{hi\perp}=z_i\bm k_{i\perp}$ and $y_{hi}=y_i$, neglecting corrections from the meson mass in this mapping. The two fragmentation fractions are integrated independently. It follows that
\begin{align}
 \bm q_{h\perp}&=\frac{z_1+z_2}{2}\bm q_\perp+(z_1-z_2)\bm P_\perp,\label{eq:fragmentation_mapping_v1}\\
 \bm P_{h\perp}&=\frac{z_1+z_2}{2}\bm P_\perp+\frac{z_1-z_2}{4}\bm q_\perp.
 \label{eq:fragmentation_mapping_p_v2}
\end{align}
Thus keeping the rapidities unchanged does not keep the pair angle unchanged. Unequal fragmentation fractions also contribute to the meson-pair total transverse momentum along $\bm P_\perp$.

Convolving Eq.~\eqref{eq:partonic_harmonics_v2} with the fragmentation functions gives the meson-pair cross section
\begin{equation}
 \begin{aligned}
 \frac{\mathrm d\sigma_{\mathrm{Xn0n}}}
 {\mathrm d^2\bm P_{h\perp}\,\mathrm d^2\bm q_{h\perp}\,\mathrm d y_{h1}\,\mathrm d y_{h2}}
 ={}&\int_0^1\frac{\mathrm d z_1}{z_1^2}\int_0^1\frac{\mathrm d z_2}{z_2^2}
 D_{Q\to H}(z_1)D_{\bar Q\to\bar H}(z_2)\\
 &\times\int_0^\infty\frac{\mathrm d b_\perp\,b_\perp}{2\pi}\,
 e^{-S_{\rm ini}(b_\perp,Q)-S_{\rm soft}(b_\perp,Q,P_\perp,M_Q)}\\
 &\times\big[C_0J_0(b_\perp q_\perp)-C_2J_2(b_\perp q_\perp)\cos(2\phi)\\
 &\hspace{13mm}+C_4J_4(b_\perp q_\perp)\cos(4\phi)+\cdots\big].
 \end{aligned}
 \label{eq:meson_cross_section_v1}
\end{equation}
We use the Peterson form to model heavy-quark hadronization~\cite{PhysRevD.27.105},
\begin{equation}
 \;{D_{Q\to H}(z)}
 =N_H\frac{z(1-z)^2}{\big[(1-z)^2+\epsilon_Q z\big]^2},
 \qquad \epsilon_c=0.05,\quad \epsilon_b=0.005.
 \label{eq:peterson_input_v1}
\end{equation}
The values of $\epsilon_c$ and $\epsilon_b$ are model parameters chosen for this calculation. The input is normalized to unity. A constant fragmentation fraction for a specified meson species cancels from the normalized observables considered here. The antiquark function follows by charge conjugation.

We present results for Pb--Pb collisions at $\sqrt{s}=5.36~\mathrm{TeV}$, with the quark masses specified in Sec.~\ref{subsec:final_state_resummation}. We take $\alpha_e=1/137$ and use the fixed coupling of Sec.~\ref{Sec. 2.2}. for the Born normalization. We consider three $P_{h\perp}$ intervals with the same rapidity cuts,
\begin{equation}
 \begin{gathered}
 P_{h\perp}\in(10,15),\ (15,20),\ (20,30)~\mathrm{GeV},\\
 0<y_{h1}<2,\qquad 0<y_{h2}<2,
 \end{gathered}
 \label{eq:meson_cuts_v1}
\end{equation}
with the two rapidities integrated independently. We restrict the gluon input to $10^{-5}\leq x_g\leq10^{-2}$ inside the fragmentation convolution. We use the directed Xn0n flux of Eq.~\eqref{eq:conditional_flux_v1}. This flux does not in general cancel from the angular moments after integration over rapidities and fragmentation fractions.

The measured angular moments are formed after fragmentation. We define

\begin{equation}
 \langle\cos(n\phi_h)\rangle
 =\frac{\displaystyle\int\frac{\mathrm d\sigma}{\mathrm d\mathcal{P.S.}}\,
 \cos(n\phi_h)\,\mathrm d\mathcal{P.S.}}
 {\displaystyle\int\frac{\mathrm d\sigma}{\mathrm d\mathcal{P.S.}}\,
 \mathrm d\mathcal{P.S.}}.
 \label{eq:meson_moments_v1}
\end{equation}
Here $\mathrm d\mathcal{P.S.}$ denotes the phase space integrated over for each observable.
The relative angle is integrated over $0\leq\phi_h<2\pi$, and the other integrals follow Eq.~\eqref{eq:meson_cuts_v1}. The moments are evaluated separately in each $P_{h\perp}$ interval.

\begin{figure}[htbp]
    \centering
    \includegraphics[width=\textwidth]{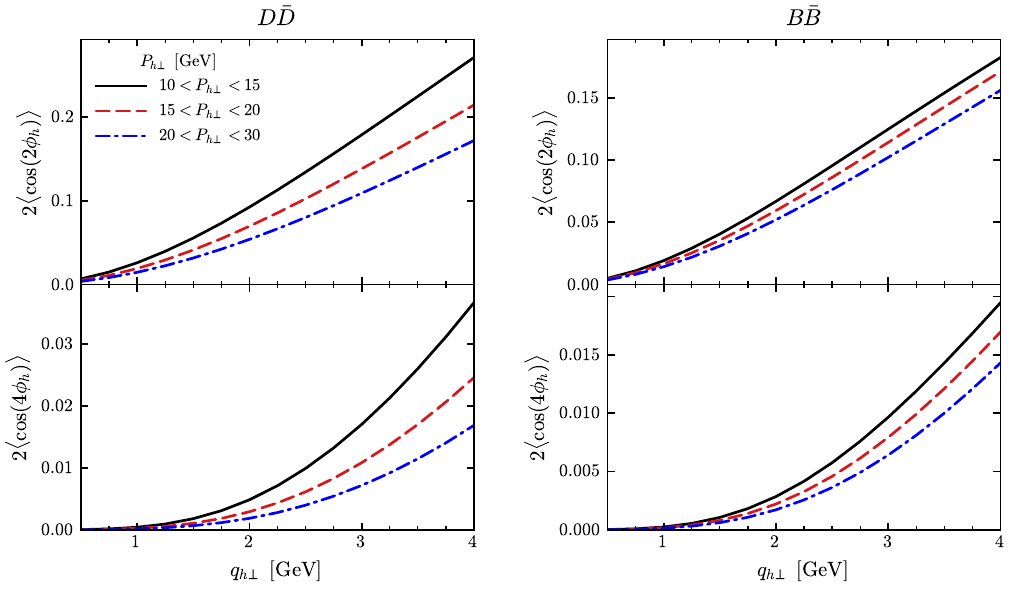}
    \caption{ Azimuthal moments for $D\bar D$ (left) and $B\bar B$ (right) production in Xn0n-selected Pb--Pb collisions at $\sqrt{s}=5.36~\mathrm{TeV}$. The two rapidities are independently integrated over $0<y_{h1},y_{h2}<2$, with $P_{h\perp}$ in the ranges 10--15, 15--20 and 20--30~GeV. The upper and lower panels show $2\langle\cos(2\phi_h)\rangle$ and $2\langle\cos(4\phi_h)\rangle$, respectively. Line styles distinguish the three momentum intervals. }
    \label{Fig. azimuthal asymmetry}
\end{figure}

Figure~\ref{Fig. azimuthal asymmetry} shows $2\langle\cos(n\phi_h)\rangle$ for $n=2,4$. Both increase with $q_{h\perp}$ over the displayed interval, and the $D\bar D$ moments are larger than the $B\bar B$ moments. At fixed $q_{h\perp}$, both moments decrease as $P_{h\perp}$ increases. The change is more pronounced for $D\bar D$. For both meson species, the $\cos(2\phi_h)$ asymmetries are much larger than those for $\cos(4\phi_h)$.

\begin{figure}[htbp]
    \centering
    \includegraphics[width=\textwidth]{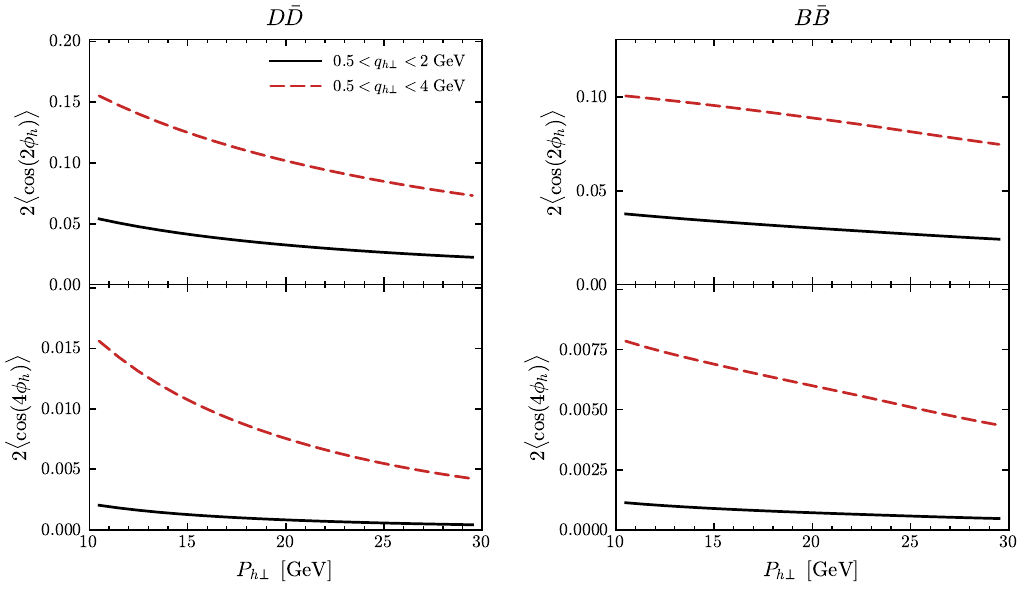}
    
    \caption{Azimuthal moments versus $P_{h\perp}$ for $D\bar D$ (left) and $B\bar B$ (right), with the same panel arrangement and other settings as in Fig.~\ref{Fig. azimuthal asymmetry}. Solid and dashed curves correspond to $0.5<q_{h\perp}<2~\mathrm{GeV}$ and $0.5<q_{h\perp}<4~\mathrm{GeV}$, respectively. Results are shown at the centers of $1~\mathrm{GeV}$ bins in $P_{h\perp}$.}
    \label{fig:meson_Ph_dependence_v1}
\end{figure}

Figure~\ref{fig:meson_Ph_dependence_v1} shows the $P_{h\perp}$ dependence after integration over two $q_{h\perp}$ intervals. Both moments decrease with $P_{h\perp}$, more strongly for $D\bar D$. The two species give similar values near the upper end of the $P_{h\perp}$ range. Extending the upper limit of $q_{h\perp}$ from 2 to 4~GeV increases the moments, especially the fourth, which is consistent with their $q_{h\perp}$ dependence in Fig.~\ref{Fig. azimuthal asymmetry}.

\begin{figure}[htbp]
    \centering
    \includegraphics[width=\textwidth]{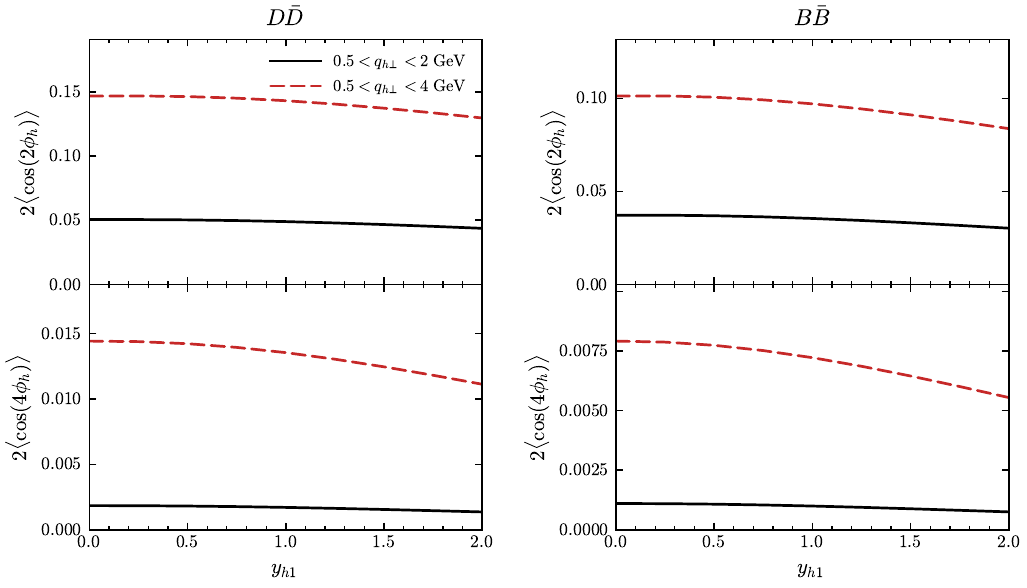}
    
    \caption{Azimuthal moments versus $y_{h1}$, integrated over $10<P_{h\perp}<30~\mathrm{GeV}$ and $0<y_{h2}<2$. The panel arrangement and other settings are as in Fig.~\ref{Fig. azimuthal asymmetry}. Solid and dashed curves correspond to $0.5<q_{h\perp}<2~\mathrm{GeV}$ and $0.5<q_{h\perp}<4~\mathrm{GeV}$, respectively.}
    \label{fig:meson_rapidity_v1}
\end{figure}

In Fig.~\ref{fig:meson_rapidity_v1}, we show the $y_{h1}$ dependence of the azimuthal asymmetries after integration over the transverse momenta and $y_{h2}$. The moments vary slowly at small $y_{h1}$ and decrease towards $y_{h1}=2$. For $0.5<q_{h\perp}<4~\mathrm{GeV}$, the second moment decreases by about $12\%$ for $D\bar D$ and $17\%$ for $B\bar B$ between $y_{h1}=0$ and 2. The relative change is larger in the fourth moment. The results as functions of $y_{h2}$ are identical: with the same rapidity cuts and charge-conjugate fragmentation functions, exchanging the two mesons leaves these even angular moments unchanged.
These moments receive contributions from gluon polarization, soft radiation, and independent fragmentation. As a check of the WW polarization contribution, we also set $h_1^{\perp g}=0$ while keeping all other inputs fixed. At $q_{h\perp}=2~\mathrm{GeV}$ in the $10<P_{h\perp}<15~\mathrm{GeV}$ interval, this decreases $2\langle\cos(2\phi_h)\rangle$ by $2.08\times10^{-4}$ for $D\bar D$ and $4.03\times10^{-3}$ for $B\bar B$. Thus most of the meson modulation in these cuts is due to soft radiation and fragmentation. A large total moment alone is not a measure of the size of the linearly polarized WW distribution.

At $q_{h\perp}=2$ and $4~\mathrm{GeV}$ in the three $P_{h\perp}$ intervals of Fig.~\ref{Fig. azimuthal asymmetry}, the Xn0n selection reduces the $D\bar D$ and $B\bar B$ differential cross sections by about $57\%$--$66\%$ compared with calculations without the neutron-selection weight, with all other inputs unchanged. The relative changes in both $\langle\cos(2\phi_h)\rangle$ and $\langle\cos(4\phi_h)\rangle$ remain below $1\%$. We keep the Xn0n selection to match the experimental event definition.

We have also performed the calculation at $5.02~\mathrm{TeV}$ and find only small changes in the angular moments.


\section{Conclusions}
\label{sec:conclusions_v2}

We have calculated $D\bar D$ and $B\bar B$ azimuthal correlations in Xn0n-selected Pb--Pb UPCs at $\sqrt{s}=5.36~\mathrm{TeV}$. The Xn0n selection is implemented through the no-neutron probability of the photon source, assuming that the hard interaction supplies the target-side neutrons. The impact-parameter-dependent photon flux is averaged over a target with a uniform transverse gluon density. The WW inputs are modeled in the CGC framework, with the $x$ dependence of their saturation scale obtained from proton rcBK evolution and nuclear scaling. CSS evolution resums logarithms from incoming-gluon radiation. We keep the full mass dependence in the Born coefficients and the one-gluon soft current, including both the angular average and the nonzero soft harmonics. The partonic cross section is then convoluted with Peterson functions to obtain the meson observables.

In the kinematic region considered, the $\cos(2\phi_h)$ and $\cos(4\phi_h)$ moments increase with $q_{h\perp}$ and decrease with $P_{h\perp}$. The second moment is much larger than the fourth. For $10<P_{h\perp}<15~\mathrm{GeV}$, the $D\bar D$ moments are larger than those for $B\bar B$; the difference becomes smaller at larger $P_{h\perp}$. Extending the $q_{h\perp}$ integration from $0.5$--$2~\mathrm{GeV}$ to $0.5$--$4~\mathrm{GeV}$ increases both moments, especially the fourth. The rapidity dependence is mild at small $y_{h1}$, with both moments decreasing towards $y_{h1}=2$. The $y_{h2}$ dependence is identical under the symmetric cuts used here. The results at $5.02~\mathrm{TeV}$ differ only slightly from those at $5.36~\mathrm{TeV}$.

For the kinematics studied, soft radiation and fragmentation give a larger contribution to the angular moments than gluon linear polarization. Extracting the polarized WW distribution from these observables therefore requires control of both effects. These predictions are obtained within the soft and correlation approximations.

\begin{acknowledgments}
We thank Jian Zhou and Shuai Yang for helpful discussions. This work was supported by the National Key R\&D Program of China under Grant No.~2024YFA1611004, the National Natural Science Foundation of China under Grant Nos.~12175117, 12321005, 12475084 and 12635003, and the Shandong Province Natural Science Foundation under Grant No. ZR2024MA012.
\end{acknowledgments}

\appendix
\section{Finite-mass soft current and angular harmonics}
\label{app:soft_mass_v1}

We give the color and kinematic steps leading to the soft factor in Sec.~\ref{subsec:final_state_resummation}. For the three colored lines in $\gamma g\to Q\bar Q$, the squared eikonal current can be written as~\cite{ZhuSunYuan:2013,Zhang:2017HQ}
\begin{equation}
 W=-k_{g\perp}^2\sum_{i,j=g,Q,\bar Q}
 \bm T_i\!\cdot\!\bm T_j\,
 \frac{v_i\cdot v_j}{(v_i\cdot k_g)(v_j\cdot k_g)},
 \qquad (v_g,v_Q,v_{\bar Q})=(p,k_1,k_2).
 \label{eq:current_colour_v1}
\end{equation}
The color operators use the all-outgoing convention, so that $\bm T_g+\bm T_Q+\bm T_{\bar Q}=0$. The signs associated with the incoming line are included in its color operator. The diagonal and interference factors are
\begin{align}
 \bm T_g^2&=C_A,\label{eq:colour_factors_v1}\\
 \bm T_Q^2&=\bm T_{\bar Q}^2=C_F,\label{eq:colour_quark_casimir_v2}\\
 \bm T_g\!\cdot\!\bm T_Q&=\bm T_g\!\cdot\!\bm T_{\bar Q}=-\frac{C_A}{2},\label{eq:colour_gluon_quark_v2}\\
 \bm T_Q\!\cdot\!\bm T_{\bar Q}&=\frac{C_A-2C_F}{2}.
 \label{eq:colour_quark_antiquark_v2}
\end{align}
Each off-diagonal term occurs twice in the sum. The incoming diagonal term vanishes because $p^2=0$, while the two massive diagonal terms give the terms proportional to $-C_FM_Q^2$ in Eq.~\eqref{eq:soft_current_mass_v1}.

At leading power in $q_\perp/P_\perp$, choose $\bm P_\perp$ along the transverse $x$ axis and boost to the frame where the mean pair rapidity is zero. With $m_T=\sqrt{P_\perp^2+M_Q^2}$ and $\Delta y=y_1-y_2$, the momenta are
\begin{align}
 p^\mu&=E_g(1,0,0,-1),\label{eq:soft_frame_v1}\\
 k_1^\mu&=\big(m_T\cosh(\Delta y/2),P_\perp,0,m_T\sinh(\Delta y/2)\big),\label{eq:soft_frame_quark_v2}\\
 k_2^\mu&=\big(m_T\cosh(\Delta y/2),-P_\perp,0,-m_T\sinh(\Delta y/2)\big),\label{eq:soft_frame_antiquark_v2}\\
 k_g^\mu&=k_{g\perp}\big(\cosh y_g,\cos(\phi_{k_gP}),\sin\phi_{k_gP},\sinh y_g\big).
 \label{eq:soft_frame_emitted_gluon_v2}
\end{align}
The incoming energy $E_g$ cancels from the current. Substituting these momenta gives
\begin{equation}
 \begin{aligned}
 &W(y_g,\phi_{k_gP})={}
 \frac{C_Ae^{\Delta y/2-y_g}}
 {\cosh(y_g-\Delta y/2)-(P_\perp/m_T)\cos(\phi_{k_gP})}\\[3pt]
 &+\frac{C_Ae^{-\Delta y/2-y_g}}
 {\cosh(y_g+\Delta y/2)+(P_\perp/m_T)\cos(\phi_{k_gP})}\\[3pt]
 &-\frac{(C_A-2C_F)\big[\cosh\Delta y+P_\perp^2/m_T^2\big]}
 {\big[\cosh(y_g-\Delta y/2)-(P_\perp/m_T)\cos(\phi_{k_gP})\big]
  \big[\cosh(y_g+\Delta y/2)+(P_\perp/m_T)\cos(\phi_{k_gP})\big]}\\[3pt]
 &-\frac{C_FM_Q^2}{m_T^2}
 \left\{\frac{1}{\big[\cosh(y_g-\Delta y/2)-(P_\perp/m_T)\cos(\phi_{k_gP})\big]^2}\right.\\[3pt]
 &\hphantom{-\frac{C_FM_Q^2}{m_T^2}}
 \left.{}+\frac{1}{\big[\cosh(y_g+\Delta y/2)+(P_\perp/m_T)\cos(\phi_{k_gP})\big]^2}\right\}.
 \end{aligned}
 \label{eq:explicit_current_v1}
\end{equation}
For $y_g\to-\infty$, $W\to4C_A$, whereas $W\to0$ for $y_g\to+\infty$. The subtraction in Eq.~\eqref{eq:soft_moments_mass_v1} can therefore be implemented as the convergent integral
\begin{equation}
 K(\phi_{k_gP})=\int_{-\infty}^{\infty}\mathrm d y_g\,
 \big[W(y_g,\phi_{k_gP})-4C_A\Theta(-y_g)\big].
 \label{eq:subtracted_current_v1}
\end{equation}
The subtracted contribution is independent of the azimuthal angle. It changes the zeroth moment but not the nonzero harmonics. The step at $y_g=0$ specifies the same subtraction as the symmetric rapidity cutoff in the pair frame; it does not introduce a mass-threshold step.

The angular average gives
\begin{equation}
 K_0=2C_A\ln\frac{m_T^2}{M_Q^2}-4C_F
 -(C_A-2C_F)\frac{1+\beta^2}{\beta}\ln\frac{1+\beta}{1-\beta},
 \qquad \beta=\sqrt{1-\frac{4M_Q^2}{Q^2}}.
 \label{eq:soft_average_check_v1}
\end{equation}
 Using $C_A-2C_F=1/C_A$, $K_0/2$ gives the coefficient in Eq.~\eqref{eq:soft_exponent_mass_v1}. The combination $\beta^{-1}\ln[(1+\beta)/(1-\beta)]$ approaches $2$ as $\beta\to0$. This relation provides a check on the mass terms and the color normalization of the numerical current.

For an even harmonic $\ell\geq2$, the radial integral in Eq.~\eqref{eq:soft_angular_mass_v1} has the form
\begin{equation}
 \int_0^Q\frac{\mathrm d k_{g\perp}}{k_{g\perp}}J_\ell(k_{g\perp}b_\perp)
 =\frac{1}{\ell}\left[1-J_0(Qb_\perp)-J_\ell(Qb_\perp)
 -2\sum_{j=2,4,\ldots}^{\ell-2}J_j(Qb_\perp)\right].
 \label{eq:bessel_integral_v1}
\end{equation}
This identity follows from the Bessel recurrence relations; the sum is empty for $\ell=2$. The integral vanishes at $b_\perp=0$ and approaches $1/\ell$ for large $Qb_\perp$. A small-argument series is used when evaluating the difference of Bessel functions would cause loss of precision.

The angular Fourier transform is
\begin{equation}
 \int_0^{2\pi}\frac{\mathrm d\phi_{bP}}{2\pi}\,
 e^{iq_\perp b_\perp\cos(\phi_{bP}-\phi)}\cos(\ell\phi_{bP})
 =(-1)^{\ell/2}J_\ell(q_\perp b_\perp)\cos(\ell\phi),
 \qquad \ell=0,2,4,\ldots .
 \label{eq:angular_transform_v1}
\end{equation}
Combining this identity with the product of the WW and soft factors gives Eqs.~\eqref{eq:meson_coefficients_v1}--\eqref{eq:meson_coefficient_c4_v2}. The higher coefficients are

\begin{equation}
 \begin{aligned}
 C_\ell&=(-1)^{\ell/2}\Big[\mathcal H_U\,x_g\tilde f_{1,\mathrm{WW}}^g(x_g,b_\perp^*)a_\ell\\
 &\qquad+\frac{\mathcal H_P}{2}\,x_g\tilde h_{1,\mathrm{WW}}^{\perp g}(x_g,b_\perp^*)
 \big(a_{\ell-2}+a_{\ell+2}\big)\Big],
 \qquad \ell=6,8,\ldots .
 \end{aligned}
 \label{eq:higher_harmonics_v1}
\end{equation}
They multiply $(-1)^{\ell/2}J_\ell(b_\perp q_\perp)\cos(\ell\phi)$ in the radial cross section. In particular, the product $\cos(2\phi_{bP})\cos(6\phi_{bP})$ supplies the $a_6$ term in $C_4$. We include soft harmonics through $\ell=14$, which generates partonic harmonics through $\ell=16$.

\section{Running coupling and threshold matching}
\label{app:evolution_numerics_v1}

For the final-state soft factor, we use
\begin{equation}
 \alpha_s(\mu)=\frac{4\pi}{\beta_0^{(n_f)}\ln(\mu^2/\Lambda_{n_f}^2)},
 \qquad \beta_0^{(n_f)}=11-\frac{2n_f}{3},
 \qquad
 n_f=\begin{cases}
 3,&\mu<M_c,\\
 4,&M_c\leq\mu<M_b,\\
 5,&\mu\geq M_b.
 \end{cases}
 \label{eq:threshold_coupling_v1}
\end{equation}
Continuity at a threshold $M_h$ fixes the next value of $\Lambda$:
\begin{equation}
 \Lambda_{n_f+1}
 =M_h\left(\frac{\Lambda_{n_f}}{M_h}\right)^{\beta_0^{(n_f)}/\beta_0^{(n_f+1)}}.
 \label{eq:lambda_matching_v1}
\end{equation}
With the parameters specified in Sec.~\ref{subsec:final_state_resummation}, Eq.~\eqref{eq:lambda_matching_v1} gives $\Lambda_4\simeq0.211~\mathrm{GeV}$ and $\Lambda_5\simeq0.161~\mathrm{GeV}$.

 Within a fixed-flavor interval, the running-coupling integral is
\begin{equation}
 \tau(\mu_2,\mu_1)
 \equiv\int_{\mu_1^2}^{\mu_2^2}\frac{\mathrm d\mu^2}{\mu^2}\frac{\alpha_s(\mu)}{2\pi}
 =\frac{2}{\beta_0^{(n_f)}}
 \ln\!\left[\frac{\ln(\mu_2^2/\Lambda_{n_f}^2)}
 {\ln(\mu_1^2/\Lambda_{n_f}^2)}\right].
 \label{eq:running_integral_v1}
\end{equation}
An interval crossing a threshold is split there, and the contributions are added. The same integral enters $S_{\rm soft}=K_0\tau(Q,\mu'_{b_*})/2$. The scale dependence of $\alpha_s$ is thus included inside the integral.

\bibliographystyle{apsrev4-2}
\bibliography{reference}

\end{document}